\documentclass[12pt,epsf,amssymb,qsymbols]{article}
\usepackage{tabularx}
\usepackage{array}
\usepackage{graphics}
\usepackage{graphicx}
\usepackage{epsfig}
\usepackage{amsmath}
\usepackage{amssymb}
\makeatletter

\usepackage{verbatim}

\newcommand{\bea}{\begin{eqnarray}}
\newcommand{\eea}{\end{eqnarray}}
\newcommand{\beq}{\begin{equation}}
\newcommand{\eeq}{\end{equation}}
\newcommand{\gev}{{\rm GeV}}

\newcommand{\pdir}{p\kern -5.2pt\raise 0.2ex\hbox {/}}
\newcommand{\vdir}{v\kern -5.75pt\raise 0.15ex\hbox {/}}
\newcommand{\kdir}{k\kern -5.75pt\raise 0.15ex\hbox {/}}
\newcommand{\epsdir}{\epsilon\kern -5.0pt\raise 0.15ex\hbox {/}}
\newcommand{\bvdir}{\bar{v}\kern -5.75pt\raise 0.15ex\hbox {/}}
\newcommand{\Ddir}{D\kern -7.75pt\raise 0.20ex\hbox {/}}
\newcommand{\ldir}{l\kern -5.0pt\raise 0.2ex\hbox{/}}
\newcommand{\varepsdir}{\varepsilon\kern -5.5pt\raise 0.15ex\hbox{/}}

\newcommand{\kkbar}{K^0-\bar K^0}
\newcommand{\bbar}{B^0-\bar B^0}

\newcommand{\bbard}{B^0_d-\bar B^0_d}

\makeatother

\begin{document}
\begin{flushright}
\begin{tabular}{l}
{\tt IJS/TP-33/02}\\
{\tt Roma-1353/02}
\end{tabular}
\end{flushright}
\begin{center}
\vskip 1.2cm\par
{\par\centering \LARGE \bf Chiral corrections and lattice QCD results}\\
\vskip 0.18cm\par
{\par\centering \LARGE \bf for  $f_{B_s}/f_{B_d}$ and $\Delta m_{B_s}/\Delta m_{B_d}$}\\
\vskip 0.75cm\par
{\par\centering \large  
\sc D.~Be\'cirevi\'c$^a$, S.~Fajfer$^{b,c}$, S.~Prelov\v{s}ek$^{b,c}$ and J.~Zupan$^b$}
{\par\centering \vskip 0.5 cm\par}
{\sl 
$^a$ Dip. di Fisica, Univ. di Roma ``La Sapienza",\\
Piazzale Aldo Moro 2, I-00185 Rome, Italy. \\                                   
\vspace{.25cm}
$^b$ J.~Stefan Institute, Jamova 39, P.O. Box 3000,\\
1001 Ljubljana, Slovenia.\\
\vspace{.25cm}
$^c$
Department of Physics, University of Ljubljana,\\
 Jadranska 19, 1000
Ljubljana,
Slovenia  }\\
{\vskip 0.25cm \par}
\end{center}

\begin{abstract}
It has been argued recently that the inclusion of the chiral logarithms in 
 extrapolation of the lattice data can shift the value of the hadronic parameter  
$\xi =  f_{B_s}\sqrt{\hat B_{B_s}}/f_{B_d}\sqrt{\hat B_{B_d}}$, from $1.16(6)$ to $1.32(10)$ and 
even higher. If true, that would considerably change the  theoretical 
estimate for the ratio of oscillation frequencies in the $B^0_s$- and $B^0_d$-systems, 
and would affect the standard CKM unitarity triangle analysis. 
In this letter we show that $f_{B_s}/f_{B_d} \approx f_K/f_\pi$, and thus the 
uncertainty due to the missing chiral logs is smaller than previously thought. 
By combining the NLO chiral expansion with the static heavy quark limit  we obtain 
$\xi = 1.22(8)$.  
\end{abstract}
\vskip 0.2cm
\setcounter{page}{1}
\setcounter{footnote}{0}
\setcounter{equation}{0}
\noindent

\renewcommand{\thefootnote}{\arabic{footnote}}

\setcounter{footnote}{0}

\vspace*{11mm}
\section{Uncertainty in $\Delta m_{B_s}/\Delta m_{B_d}$ 
comes from $f_{B_s}/f_{B_d}$}

The $\kkbar$ and $\bbar$ mixing amplitudes are the main ingredients in the 
standard unitarity triangle analysis~\cite{UTA}. They overconstrain the apex of the CKM triangle 
in the $(\bar \rho, \bar \eta)$-plane. Confronting such an overconstrained triangle with 
the directly measured $\sin(2\beta)$~\cite{babar+belle} allows for the consistency check 
from which we hope to either see the effects of the non-Standard Model physics, or simply to 
confirm the validity of Standard Model at the accessible accuracy~\cite{nir}.
One of the essential constraints is provided by the ratio of oscillation frequencies 
\bea \label{circle}
{\Delta m_{B_d}\over \Delta m_{B_s}} = \left|\frac{V_{td}}{V_{ts}}\right|^2 
\frac{m_{B_d}}{m_{B_s}}\; \frac{1}{\xi^2}\simeq \lambda^2 \left[ (1-\bar{\rho})^2+\bar{\eta}^2\right] 
\frac{m_{B_d}}{m_{B_s}}\; \frac{1}{\xi^2}+{\cal O}(\lambda^4) \;,
\eea
where $m_{B_{s(d)}}$ is the mass of $B^0_{s(d)}$-meson, while  $V_{ts}$ and $V_{td}$ are the 
CKM matrix elements. The mass difference $\Delta m_{B_d}$, which measures the 
oscillation frequency in the $\bbard$ system, has been determined accurately  in the 
experiments, $\Delta m_{B_d}=0.503 \pm 0.006 \;\text{ps}^{-1}$~\cite{stocchi}. 
On the other hand, $\Delta m_{B_s}$ is currently only bounded from below, 
$\Delta m_{B_s}> 14.4 \;\text{ps}^{-1}$~\cite{stocchi}, but will hopefully 
 be measured soon at Tevatron~\cite{tevatron}. 
On the r.h.s. of eq.~\eqref{circle}, one has the hadronic parameter
\bea
\xi = {f_{B_s} \sqrt{\hat B_{B_s}} \over f_{B_d} \sqrt{\hat B_{B_d}} } \;,
\eea
with $f_{B_q}$ and  $\hat B_{B_q}$ being the decay constant and the so called ``bag"-parameter, respectively. 
The value of the parameter $\xi$ is expected to be determined from the QCD simulations on the lattice 
with a good accuracy since many uncertainties should cancel in the ratio.

Recently, however, it has been 
argued that the chiral logarithms might be a source of a large, yet unaccounted for, uncertainty in 
$\xi$~\cite{kronfeld}. The reason for this lies in the fact that the masses of light quarks that are  directly 
accessible from the lattice studies are those around the strange quark,  
$m_s/2 \lesssim m_q \lesssim 3 m_s/2$. In this range $f_{B_q}$ and  $\hat B_{B_q}$, computed on the lattice, 
exhibit a linear behaviour under the variation of $m_q$.  This linear dependence is then 
extrapolated in $m_q \to m_d$, to reach  $f_{B_d}$ and  $\hat B_{B_d}$.
The claim of ref.~\cite{kronfeld} is that the inclusion of the chiral logarithms in the 
mentioned extrapolation produces a shift from the ``standard" value $f_{B_{s}}/f_{B_{d}} = 1.16(5)$, to 
$1.32(8)$. In ref.~\cite{yamada} the systematic error due to the  shift is estimated to be even larger, 
$+0.24$. In the following we revise the problem 
and argue that the shift is indeed present but it is actually {\it smaller} than previously claimed.
We will show that $f_{B_{s}}/f_{B_{d}} \approx f_K/f_\pi$.

Finally, from refs.~\cite{kronfeld,grinstein} it is known that, w.r.t. the result of the linear 
extrapolation in the light  quark mass, the chiral logs produce a tiny (insignificant) shift 
in the ratio of ``bag"-parameters, $B_{B_{s}}/B_{B_{d}}$. 
This latter ratio is known to be very consistent with unity~\cite{B-params}, and therefore 
is not the source of any additional uncertainty in the determination of $\xi$.

\section{Ratios $f_{B_s}/f_{B_d}$ and $f_{K}/f_{\pi}$ from the ChPT \label{ratiosRBRP}}

The SU(3) light flavour breaking effects in the ratios of the decay constants have been calculated  
at the 1-loop level of the chiral perturbation theory (ChPT), resulting in the following expressions~\cite{gasser,cho}
\bea \label{rB}
R^{\text{ChPT}}_{f_B} \equiv { \Phi_{B_s} \over \Phi_{B_d}  } = 1 + {1 + 3 g^2\over 4 (4 \pi f)^2} \left[
3 I_1(m_\pi^2) - 2  I_1(m_K^2) - I_1(m_\eta^2) \right] + {8 K\over f^2} (m_K^2 - m_\pi^2)  \,,
\eea
\bea \label{rP}
R^{\text{ChPT}}_{f_\pi} \equiv {f_{K}\over f_{\pi}} = 1 + {1 \over 4 (4 \pi f)^2} \left[
5 I_1(m_\pi^2) - 2  I_1(m_K^2) - 3 I_1(m_\eta^2) \right] + {8 L_5\over f^2} (m_K^2 - m_\pi^2)  \,,
\eea
with the convention $f=130$~MeV, and the standard notation $\Phi_{B_q} = f_{B_q}\sqrt{m_{B_q}}$. 
The counter-term $L_5$ is defined in \cite{gasser}, while $K$ stands for a sum of counter-terms discussed in \cite{cho}.
Note that these counter-terms provide the corrections to $R^{\text{ChPT}}_{f_\pi,f_B}$ 
 linear in quark mass and that they eliminate the renormalization scale dependence  arising from the 
chiral loop integral
\bea
I_1(m^2) = m^2 \log(m^2/\mu^2)\;.
\eea
The renormalization scheme of ref.~\cite{gasser} has been used.
The only new parameter accompanying the chiral logarithm in eq.~(\ref{rB}) is $g$, 
the coupling of the lowest lying spin doublet of heavy mesons to a soft pion. Its experimental 
value has been recently established in the charm sector, namely $g_c=0.59(8)$~\cite{cleo}. 
While a recent (quenched) lattice study suggests that the $g$-coupling does not change 
when increasing the heavy quark mass ($g_b\approx g_c$)~\cite{orsay},  
the light cone QCD sum rule calculation predicts the suppression by about $25\%
$ (i.e. $g_b/g_c \approx 0.75$)~\cite{khodjamirian}. We will take the average of 
the two predictions and add the difference in the systematic uncertainty, i.e.  
$g \equiv g_b=0.52(7)(7)$. 
It is worth mentioning that the use of the tree level values for $g$ and $f$ is consistent 
with the use of the ratios~\eqref{rB} and \eqref{rP}, derived at next-to-leading order in 
the chiral expansion.

Eqs.~(\ref{rB}) and~(\ref{rP}) have been obtained in refs.~\cite{cho} and~\cite{gasser}, 
respectively. Notice that in the former case we neglect the ${\cal O}(1/m_b)$-corrections, 
which are anyway expected to cancel in the ratio~(\ref{rB}) to a large extent \cite{cho}. 
Throughout this letter we assume the isospin symmetry $m_d=m_u\equiv m_q$, and use the 
Gell-Mann--Oakes--Renner and Gell-Mann--Okubo formulas to write 
\bea \label{rmass}
m_\pi^2 = 2 B_0 m_s r\,,\quad m_K^2 = 2 B_0  m_s { r+1  \over 2}\,,\quad 
m_\eta^2 = 2 B_0 m_s { r+2 \over 3}\,, 
\eea
where $r=m_q/m_s$, and $B_0$ is the common factor related to the quark condensate, 
$B_0 = - 2 \langle \bar q q \rangle/f^2$. When varying the ratio $r$, we will always keep the 
strange quark mass fixed to its physical value. We can then use the physical 
kaon and pion masses to get $ 2 B_0 m_s$ = $2 (m_K^{phys})^2 - (m_\pi^{phys})^2 $ = $0.468\ \gev^2$.

\begin{figure}
\vspace*{-0.8cm}
\begin{center}
\begin{tabular}{@{\hspace{-0.25cm}}c}
\epsfxsize9.2cm\epsffile{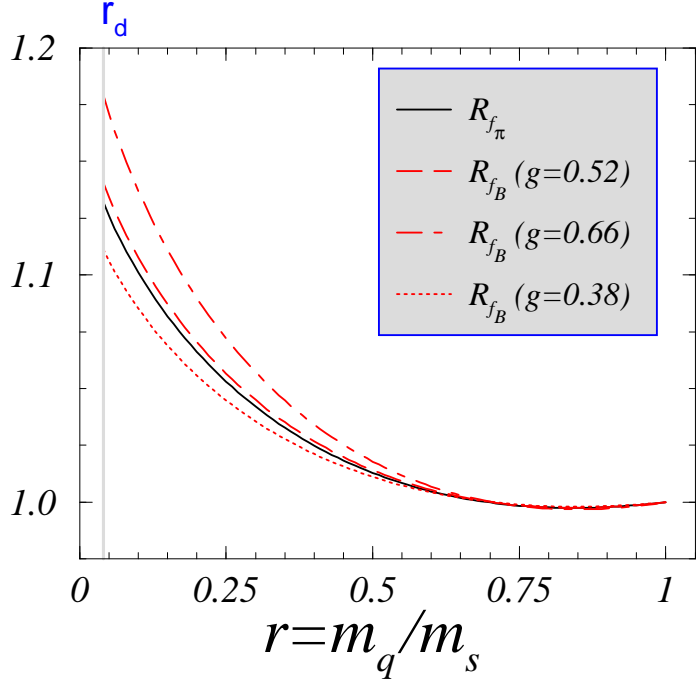}    \\
\end{tabular}
\vspace*{-.5cm}
\caption{\label{figLog}{\footnotesize The chiral logarithmic contributions to $R_{f_\pi}$ and 
$R_{f_B}$  for three different values of $g$ spanning the range $g=0.52(7)(7)$, discussed in the text. 
Plotted are eqs. \eqref{rB} and \eqref{rP} with $K=L_5=0$ and $\mu = 1$~GeV. 
Variation of $r=m_q/m_s$ is made by keeping $m_s$ fixed to its physical value.} }
\end{center}
\end{figure}

It is important to note that even though the chiral logarithms in eqs.~\eqref{rB} and 
\eqref{rP} enter with different coefficients, they are numerically very similar. 
To illustrate this point we set $K=L_5=0$ and plot in fig.~\ref{figLog} only the chiral 
log contributions to $R_{f_B}$ and $R_{f_\pi}$. 
Indeed, we see that for the central value of the parameter $g$, the chiral log contributions 
to $R_{f_B}$ and $R_{f_\pi}$ nearly coincide, whereas the relative difference at the physical point  
$r\simeq 0.04$ is  below $5\% 
$ level when $g$ is varied in the range $g \in (0.35, 0.7)$. Note that the near coincidence 
of chiral logarithmic terms in $R_{f_B}$ and $R_{f_\pi}$ is not spoiled by the change of 
the renormalization scale $\mu$.

From fig.~\ref{figLog} we also observe the well known fact that the chiral logs 
are not large enough to reproduce the experimental value $(f_K/f_\pi)_\text{exp.} = 1.22(1)$~\cite{pdg}. 
The mismatch is patched by fixing the coupling $L_5(\mu = 1\ \gev)$ to $(8.7\pm 2.5)\times 10^{-4}$ 
from eq. \eqref{rP}. The remaining entry in eq.~\eqref{rB}, the value of $K$,  
will be discussed later on, at the end of section~\ref{double}.

\section{Chiral extrapolation of $f_{B_s}/f_{B_q}$}
Contrary to the ChPT, which is valid for very small quark masses, the lattice QCD 
provides predictions for not so light quarks corresponding to $1/2 \lesssim r \lesssim 3/2$. 
In other words, $f_{B_s}$ ($r=r_s=1$) is accessed directly from the lattice studies, while for 
$f_{B_d}$ an extrapolation in $r\to r_d$ is needed ($r_d\simeq 0.04$~\cite{leutwyler}). 
That extrapolation is typically made linearly as 
\bea \label{linear}
R^{\rm latt}_{f_B} = 1 + \alpha \left( 1 - r \right)\;,
\eea
with $\alpha=0.16(5)$~\cite{yamada,sinead}. The form~\eqref{linear} does not include the chiral 
logarithmic terms shown in eq.~\eqref{rB}, which would increase the result of extrapolation, 
namely $f_{B_s}/f_{B_d}$. It is, however, not clear at what value of $r$ the leading chiral logs 
need to be included, i.e. we do not know how light the ``pions" should be so that the leading 
chiral logs remain uncompensated by the higher order terms in the chiral expansion. 

\begin{figure}[h]
\vspace*{-0.85cm}
\begin{center}
\begin{tabular}{@{\hspace{-.95cm}}c}
\epsfxsize11.4cm\epsffile{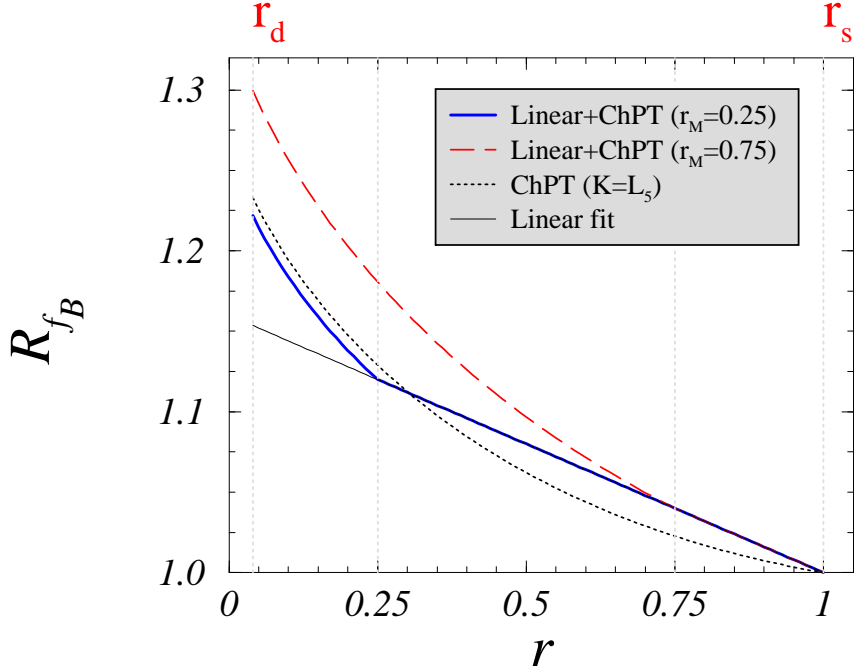}    \\
\end{tabular}
\vspace*{-.85cm}
\caption{\label{fig1}{\footnotesize Chiral extrapolation of the ratio 
$R_{f_B}=f_{B_s}/f_{B_q}$, in which the chiral logarithms~\eqref{rB} are 
included  from $r_M=0.25$ and $0.75$ with $K$ fixed from~\eqref{Kfix} as in 
ref.~\cite{kronfeld}. Note that the variation of $r=m_q/m_s$ is made by keeping $m_s$ 
fixed to its physical value. The illustration is provided for $g=0.52$. 
We also show $R_{f_B}=f_{B_s}/f_{B_q}$  with $K$ fixed as $K=L_5$ (dotted line).} }
\end{center}
\end{figure}

In the approach adopted in ref.~\cite{kronfeld} the NLO prediction of ChPT~\eqref{rB} 
is assumed to be valid up to $r=1$, while the counter-term $K$ is fixed from the requirement
\beq \label{Kfix}
\bigl. R^{\rm ChPT}_{f_B}\bigr|_{r=r_M}= \left.R_{f_B}^{\rm latt}\right|_{r=r_M}  \;,
\eeq
at some intermediate point $r_M$. We slightly modify that approach by actually separating 2 
regions: (i) above $r_M$, where the extrapolation of the linearly fitted 
lattice data~(\ref{linear}) is a good approximation; (ii) below $r_M$, where we use  the 
ChPT formula~\eqref{rB} with $K$ fixed from the condition~\eqref{Kfix}, and go  
down to the physical point $r_d$. The effect is shown in fig.~\ref{fig1} where we 
plot two curves, corresponding to $r_M = 0.25$, and $0.75$~\footnote{For an easier 
orientation we convert 
$r_M$ by using $m_M^2=2 B_0 m_s r_M$ ($2 B_0 m_s = 0.468\ \gev^2$) to obtain 
$m_M\approx 330$ and $580$~MeV, respectively.}.
From fig.~\ref{fig1}  we see that farther away from $r_d$ lies the point $r_M$, 
 the larger is the shift in $R_{f_B}$. By using $g=0.52$, 
we obtain~\footnote{Had we used the approach similar to the one in ref.~\cite{bpi} with  
{\it both} the size and the derivative of $R_{f_B}$  matched to the lattice data at $r_M$, 
the corresponding shifts would be smaller than the ones presented in~\eqref{shifts}. 
Note that this procedure amounts to introducing an additional constant term $K'$ in 
\eqref{rB} (arising from NNLO contributions) with both $K$ and $K'$ then fixed from lattice.}
\bea
R_{f_B} =   1.22(6)_{r_M=0.25} ,\quad 1.30(7)_{r_M=0.75}\,,\label{shifts}
\eea
to be compared with the result of the linear extrapolation $R_{f_B} =1.16(5)$.
To estimate this shift, the authors of ref.~\cite{kronfeld} take $r_M=0.75\pm 0.25$.
Although the relevance of the leading chiral logs for the matching scale at $m_M \simeq 
( 580\pm 100)$~MeV may be doubted, at this point there is no legitimate argument 
to exclude the possibility that the central value for $R_{f_B}$ gets as high as $1.3$.

\section{Double ratio $(f_{B_s}/f_{B_d})/(f_K/f_\pi)$}\label{double}

To get around the above ambiguity we construct the double ratio~\cite{hldc}
\bea\label{rr}
R= {\Phi_{B_s}/ \Phi_{B_d} \over f_{K}/ f_{\pi} } &=& 
1 - {1\over 4 (4 \pi f)^2} \biggl[ (2 -9 g^2)  
 I_1(m_\pi^2) + 6 g^2  I_1(m_K^2) - (2 -3 g^2) I_1(m_\eta^2) \biggr] \nonumber \\
&& \hspace*{24mm}+ {8 (K-L_5)\over f^2} (m_K^2 - m_\pi^2) \;. 
\eea
As shown at the end of section~\ref{ratiosRBRP}, the chiral log contributions in $R=R_{f_B}/R_{f_\pi}$ 
almost completely cancel so that the double ratio $R$, to a very good approximation, 
depends on $r$ only linearly. One can then envisage a future lattice analysis in which 
the chiral  extrapolation of $R$ is made linearly and then 
the result for  $f_{B_s}/ f_{B_d}$ is deduced by using the experimental value 
$(f_K/f_\pi)_{\rm exp.} = 1.22(1)$.

In this letter, we will use the formula~\eqref{rr} to estimate the value of $f_{B_s}/f_{B_d}$. 
To do so we have to fix the couplings $L_5$ and $K$. We already mentioned that  
 $L_5=(8.7\pm 2.5)\times 10^{-4}$ (at $\mu = 1$~GeV). However, the coupling $K$ is unknown. 
To get its value, but only relative to $L_5$, we may rely on the experience with the QCD 
simulations on the lattice in which the heavy-light ($B_q$) and light-light ($P_{qq}$) 
pseudoscalar meson decay constants are fitted well to the  linear forms  
\bea
f_{B_q} = a_0 + a_1 m_{P_{qq}}^2 \;, \qquad {\rm and} \qquad  f_{P_{qq^\prime }} = b_0 + b_1 m_{P_{qq^\prime }}^2 \;.
\eea
By using eq.~\eqref{rmass}, these forms can be rewritten as
\bea \label{fitt}
R_{f_B} \simeq 1 + \underbrace{{a_1 \over a_0} 2 B_0 m_s}_{\displaystyle \alpha} (1 - r) \;, 
\qquad {\rm and} \qquad  
R_{f_\pi} \simeq 1 + \underbrace{{b_1 \over 2 b_0} 2 B_0 m_s}_{\displaystyle \beta} (1 - r) \;. \label{alphabeta}
\eea
The values of parameters $\alpha$ and $\beta$ can be obtained from the linearly fitted lattice 
data for $R_{f_B}$ and $R_{f_\pi}$, which at $r=0$ simply read, $R_{f_B}=1+\alpha$  and $R_{f_\pi}=1+\beta$. 
From the recent compilation of  unquenched calculations (with $n_f=2$) of the heavy-light decay constants, 
which are obtained through the linear extrapolation, we take $f_{B_s}/f_{B_d}=1.16(5)$~\cite{yamada,sinead}. 
The MILC collaboration has recently verified that the ratio $f_{B_s}/f_{B_d}$ remains stable when the 
number of dynamical flavors $n_f$ and/or the values of their masses are changed~\cite{Bernard:2002ep}. 
We therefore take $\alpha = 0.16(5)$. From an extensive study of the light 
hadrons with two flavors of dynamical quarks by the JLQCD and the CP-PACS collaborations~\cite{AliKhan:2001tx}, 
one has $f_K/f_\pi=1.16(2)$, i.e.  $\beta=0.16(2)$.~\footnote{This result corresponds to the average of the results 
obtained by using $K$- and $\phi$- inputs in the recent paper by the JLQCD Collaboration in which the results of the 
improved unquenched study were reported.}  
In other words, $\alpha/\beta \approx 1$. That conclusion is also in agreement with 
the observation made by the MILC collaboration that the slope $b_1$ is about 
$30\%
$ larger than  the slope $a_1$~\cite{claude}~\footnote{We thank Claude Bernard for this information.}. 
Moreover, by fitting, linearly in $r$, the unquenched data for both the light-light and the heavy-light decay 
constants obtained by the JLQCD collaboration and presented in ref.~\cite{kaneko}, one arrives to the same 
conclusion, namely $\alpha/\beta \approx 1$.

This information can be used to estimate the value of $K-L_5$ by matching at some intermediate point $r_M$ 
the  chiral expression for $R$ given in eq. \eqref{rr} to the linear fits of lattice data given in eq.~\eqref{alphabeta}
\bea
\left. R\right|_{r=r_M}=\left. \left(\frac{R_{f_B}^{\text{latt}}}{R_{f_\pi}^{\text{latt}}}\right)\right|_{r=r_M}. \label{match}
\eea
In further discussion we will take $r_M$ to be in the range $r_M\in (0.5,1)$, where lattice simulations are performed.  
From requirement \eqref{match} it follows trivially
\bea
\frac{8 (m_K^{\text{phys}})^2}{f^2} (K-L_5)=(\alpha-\beta)- \frac{1}{1-r_M}\left .\delta R^{\text{loop}}\right|_{r=r_M} ,\label{K-L_5}
\eea
where $\delta R^{\text{loop}}$ is  the chiral log correction to the $R$ ratio (i.e. the first correction term in 
eq.~\eqref{rr}), while $\alpha$ and $\beta$ are the slopes of the fits to the lattice data points as defined  
in eq.~\eqref{alphabeta}. Finally, the expression for the double ratio \eqref{rr} is 
\bea
R=1+\delta R^{\text{loop}}+\left((\alpha-\beta)-\frac{1}{1-r_M} \left .\delta R^{\text{loop}}\right|_{r=r_M}\right) (1-r)\label{Rfrommatch}
\eea
As we already mentioned, the available lattice data suggest $\alpha\approx \beta$. Furthermore, the size of chiral corrections to the 
matching condition,  $\left .\delta R^{\text{loop}}\right|_{r=r_M}/(1-r_M)$, is below $1\%
$ for $r_M\in (0.5,1)$ and $g\in(0.35,0.7)$. As such it is much smaller then the present errors on the value of $\alpha-\beta$ 
and can be safely neglected in the following.  

As a first estimate we can thus set the last term in eq.~\eqref{Rfrommatch} to zero. 
At the physical point $r=r_d=0.04$, and by using $g =0.52(7)(7)$, we get 
\bea
R=1.011 \pm 0.024 \;.
\eea 
where the errors reflect the variation of the $g$-parameter only. From this result, 
we conclude that $f_{B_s}/f_{B_d} \approx f_K/f_\pi$.

\section{Our estimate of $f_{B_s}/f_{B_d}$ and $\xi$}

Finally, to make the numerical estimate we take 
\bea\label{input}
 f=0.13~\gev\,,\quad g=0.52(7)(7)\,, \quad K/L_5 = 1.0 \pm 0.6\,, \quad  L_5=(8.7\pm 2.5)\times 10^{-4}\,,
\eea
where the first error in $g$ is considered to be gaussian (experimental) and all the others to be flat. 
The error in $(f_K/f_\pi)_{\rm exp.} = 1.22(1)$ is also gaussian. 
Notice in particular that the value of  $K/L_5$ ratio follows from the present value of $\alpha-\beta=0.00\pm 0.06$, 
discussed in the previous section,  through the use of eq.~\eqref{K-L_5}. That uncertainty may easily be reduced 
by the direct lattice calculation of the ratio $R$.

\begin{figure}[h!!]
\begin{center}
\begin{tabular}{@{\hspace{-.95cm}}c}
\epsfxsize10.0cm\epsffile{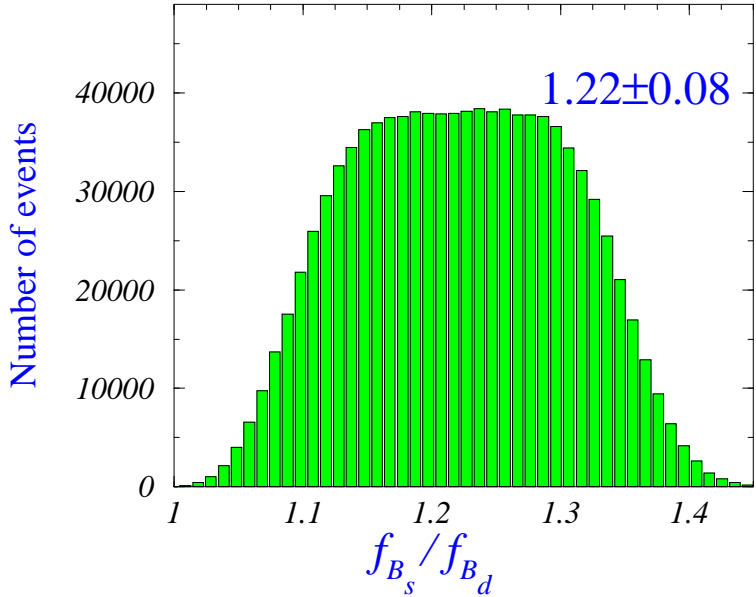}    \\
\end{tabular}
\vspace*{-.15cm}
\caption{\label{fig3}{\footnotesize Probability distribution function for the values 
of the SU(3) breaking ratio $f_{B_s}/f_{B_d}$, as obtained by using the input values 
discussed in the text (see eq.~\eqref{input}). 
} }
\end{center}
\end{figure}

With these input values and by using eq.~\eqref{rr}, we produced 
the  histogram of $10^6$ Monte Carlo events for 
$f_{B_s}/f_{B_d} = R \sqrt{m_{B_d}/m_{B_s}}\times (f_K/f_\pi)_\text{exp.}$, 
which is shown in fig.~\ref{fig3}.
We finally obtain 
\bea \label{RES}
f_{B_s}/f_{B_d} = 1.22 \pm 0.08\;,
\eea
which is larger than the commonly quoted values, yet it is considerably smaller than the results obtained 
in refs.~\cite{kronfeld,yamada}. 

To make the discussion complete and get an estimate of the parameter $\xi$, 
we will also use the result of ref.~\cite{grinstein}:  
\bea \label{bb}
 {\hat B_{B_s} \over \hat B_{B_d}} =
1 + {1 -3 g^2 \over 2 (4 \pi f)^2} \left[   
 I_1(m_\pi^2) -  I_1(m_\eta^2) \right] + {16 \widetilde K\over f^2} (m_K^2 - m_\pi^2)\;,\label{Bfac}
\eea
where $\widetilde{K}$ denotes the sum of corresponding counter-terms. Combining eq.~\eqref{Bfac} with 
eq.~\eqref{rr} gives
\bea \label{cal}
{\cal R}_\xi= {\xi \sqrt{m_{B_s}/m_{B_d}}\over f_K/f_\pi} &=& 1 - {1 \over 4 (4 \pi f)^2} \left[
(1-6 g^2) I_1(m_\pi^2) + 6 g^2  I_1(m_K^2) - I_1(m_\eta^2) \right]\nonumber \\
&& \quad + {8 (K+ \widetilde K - L_5)\over f^2} (m_K^2 - m_\pi^2)\,.
\eea
As mentioned at the beginning of this letter, the results of several lattice studies suggest  
that $\widetilde K$ can be safely set to zero~\cite{B-params}. With the same input values, given in 
eq.~\eqref{input}, we compute ${\cal R}_\xi \sqrt{m_{B_d}/m_{B_s}}\times (f_K/f_\pi)_{\rm exp.}$ 
from eq.~\eqref{cal}, and obtain
\bea
\xi = 1.22 \pm 0.08\;.
\eea

It is important to stress that our results for  $f_{B_s}/f_{B_d}$ and $\xi$ are obtained 
by neglecting the ${\cal O}(1/m_b^{n>0})$ corrections that survive the cancellation in 
the ratio~\eqref{rB}.  One may worry about the possibility that those corrections may 
further increase our results. The experience with lattice QCD and with the QCD sum rule (QSR)
calculations suggests that the SU(3) breaking ratio of the decay constants {\it is not} 
decreasing when increasing  the heavy meson mass. Indeed, from ref.~\cite{sinead} 
we read the lattice QCD estimates, $f_{B_s}/f_{B} = 1.16(5)$,  $f_{D_s}/f_{D} = 1.12(4)$, while 
from the last paper of ref.~\cite{B-params} we see that the QSR calculation gives,  
$f_{B_s}/f_{B} = 1.16(5)$,  $f_{D_s}/f_{D} = 1.15(4)$. Therefore the uncancelled ${\cal O}(1/m_b^n)$ 
corrections are likely to shift our result~\eqref{RES}  towards the {\it smaller} values rather than  
the larger ones.

Notice that a  conclusion similar to ours' has been also reached in the recent reanalysis of the lattice data 
produced by the MILC collaboration~\cite{claude}. The authors show that the consistent inclusion of 
the chiral logarithms in all the chiral extrapolations involved in their data analysis, produces a shift 
of their central result, $f_{B_s}/f_{B_d} = 1.16(1)(2)(2)$, by only $+0.04$, which is much smaller 
than $+0.24$, as claimed in ref.~\cite{yamada}, or  $+0.16$, as argued in ref.~\cite{kronfeld}.
Moreover, by actually shifting their result, one gets $f_{B_s}/f_{B_d} = 1.20(1)(2)(2)$, which is 
completely consistent with our main claim, namely $f_{B_s}/f_{B_d} \approx f_K/f_\pi$. 

Finally we mention, that instead of the double ratio  $(f_{B_s}/f_{B_d})/(f_K/f_\pi)$  discussed in 
the present letter, the double ratio of the {\it heavy meson} decay constants $R_1=(f_{B_s}/f_{B_d})/(f_{D_s}/f_{D_d})$ 
could be used for the determination of $(f_{B_s}/f_{B_d})$ and correspondingly of $\xi$  as suggested by 
Grinstein in~\cite{Grinstein:1993ys}. The ratio of charm meson decay constants $(f_{D_s}/f_{D_d})$ 
is expected to be measured by CLEO-c, while preliminary $N_f=2$ dynamical lattice calculations of 
$R_1$  are already available~\cite{yamada}. 

\section{Conclusions}
In this letter we have discussed the effect of the inclusion of the chiral logarithms on the lattice
determination of the ratio $f_{B_s}/f_{B_d}$. We have shown that in the double ratio 
\beq\nonumber
R={f_{B_s}/ f_{B_d} \over f_{K}/ f_{\pi} }\;,
\eeq
the chiral logarithms almost completely cancel because of the fact that the numerical value of the  
coupling  $g$ is close to $\sim 0.5$.  To a very good approximation the double ratio $R$ then 
depends {\it linearly} on the light quark mass all the way down to the chiral limit. 

We have also made a numerical estimate of the hadronic parameter $\xi$, by using the expressions 
derived in the ChPT and by fixing the unknown couplings from the available lattice results. 
Namely, we obtain
\beq
\xi=1.22\pm0.08\,.\nonumber
\eeq

\vskip 11mm 

\section*{Acknowledgement} 
It is a pleasure to thank A.~Buras, L.~Lellouch, V.~Lubicz, G.~Martinelli 
and C.~Sachrajda for the comments on the manuscript. We also kindly acknowledge 
the correspondance with T.~Kaneko concerning the details on the results of ref.~\cite{AliKhan:2001tx}. 
The research of 
S.F., P.S., and J.Z. was supported in part by the Ministry of Education, 
Science and Sport of the Republic of Slovenia.


\end{document}